\documentclass[aps,showpacs,floatfix]{revtex4}
\usepackage{graphics}
\newcommand{\bea}{\begin{eqnarray}}
\newcommand{\eea}{\end{eqnarray}}
\newcommand{\be}{\begin{equation}}
\newcommand{\ee}{\end{equation}}

\def\be{\begin{eqnarray}}
\def\ee{\end{eqnarray}}
\def\bd{\begin{displaymath}}
\def\ed{\end{displaymath}}

\def\etal{{\em et al}}

\def\NP{Nucl. Phys. }
\def\PR{Phys. Rev. }
\def\PRL{Phys. Rev. Lett. }
\def\PL{Phys. Lett. }

\begin{document}
\title{$\alpha$-Decay Lifetime in Superheavy Nuclei With $A>282$}
\author{Madhubrata Bhattacharya}
\author{G. Gangopadhyay}
\email{ggphy@caluniv.ac.in}
\affiliation{Department of Physics, University of Calcutta\\
92, Acharya Prafulla Chandra Road, Kolkata-700 009, India}
\begin{abstract}
Nuclei with $A>282$ have been studied in the Relativistic Mean Field approach 
using the force FSU Gold and a zero range pairing interaction. The 
Euler-Lagrange equations have been solved in the co-ordinate space. Alpha 
nucleus potential has been constructed with the DDM3Y1 interaction, which has 
an exponential density dependence,  in the 
double folding model using the nucleon densities in the daughter nucleus and the 
$\alpha$ particle. Half lives of $\alpha$ decay have been calculated for tunneling of 
the $\alpha$ particle through the potential barrier in the WKB approximation 
and assuming a constant 
preformation probability. The resulting values agree well with experimental 
measurements.

\end{abstract}
\maketitle


Study of $\alpha$ decay in superheavy nuclei (SHN) presents a unique opportunity 
of probing the nuclear density in this mass region where other common methods 
such as scattering  are not possible yet. Alpha decay is known to take 
place through tunneling of the potential barrier by the $\alpha$ particle.
The barrier itself depends on the density profile of the daughter
nucleus. Thus, $\alpha$ decay lifetime may provide a stringent test of the ability 
of nuclear structure theories to predict the nuclear density.


In the present work, we have followed the microscopic Super-Asymmetric Fission 
Model (SAFM) which uses WKB approximation to  calculate the tunneling 
probability. The potential between the $\alpha$ particle and the daughter 
nucleus has been obtained in the double folding model by folding the proton and
neutron densities 
in the $\alpha$ particle and the daughter nucleus with some suitable interaction. 
Usually the densities are obtained from phenomenological description. However,
in the present work we utilize the microscopic densities obtained from 
Relativistic Mean Field (RMF) calculation. The shape of the barrier
is known to be sensitively dependent on the density. For example, the height
of the fission barrier in very heavy nuclei is known be to change with the 
nucleon density\cite{nucden}. Avrigeanu {\em et al}. \cite{alphascat} have investigated the 
effect of different phenomenological densities in the $\alpha$ nucleus
scattering and have found that some densities are more suited to describe 
$\alpha$ nucleus interaction.
In the superheavy region, nucleon
densities are not experimentally known and theoretical densities 
may be used as a substitute.

We study the even-even and odd mass nuclei with $A>282$ in the present report. 
A large number of calculations of half life of SHN based on the SAFM is 
available in the literature
and we cite only a few recent ones. A number of nuclei in this mass 
region 
has been systematically studied
using a phenomenological density profile and effective interaction by 
Mohr\cite{Mohr1} and Roy Chowdhury \etal\cite{Roy}. Gambhir
\etal.\cite{gambhir} have also calculated the half life values of superheavy nuclei 
with densities obtained from RMF calculation with the relativistic force 
NL3\cite{NL3}
and the density dependent interaction DDM3Y in the double 
folding approach. In the present brief report, we use a new Lagrangian density 
to calculate the density of the nuclei. We also employ an interaction
with an exponential density dependence which  reproduces the low energy
$\alpha$ scattering data very well. In contrast to most earlier works in RMF, we 
perform our calculation in the co-ordinate space as obtaining exact values for
the density as a function of radius is of great importance in constructing the
barrier.

RMF is now a standard approach in low energy nuclear structure.
It can describe various features
of stable and exotic nuclei including ground state binding energy, shape,
size, properties of excited states, single particle structure, neutron halo, 
etc\cite{Ring}.
In nuclei far away from the
stability valley, the single particle level structure undergoes modifications
in which the spin-orbit splitting plays an important role.
In exotic nuclei, it is often difficult experimentally to obtain information 
about these changes.
For example in the nuclei studied in this work, almost
nothing is known experimentally about the single particle levels.
Being based on the Dirac Lagrangian density, RMF is particularly suited to
investigate these nuclei because it naturally incorporates the
spin degrees of freedom. 
There  are different variations of the Lagrangian density and also
a number of different parameterizations in RMF. Recently, a new Lagrangian
density has been proposed\cite{prl} which involves
self-coupling of the vector-isoscalar meson as well as coupling between the
vector-isoscalar meson and the vector-isovector meson. The corresponding
parameter set is called FSU Gold\cite{prl}. This Lagrangian density has earlier 
been employed to obtain the proton nucleus interaction to successfully 
calculate the half life for proton radioactivity\cite{plb} and cluster
radioactivity\cite{clus}. In this work also, 
we have employed this force. 

In the conventional RMF+BCS approach, the Euler-Lagrange
equations are solved under the assumptions of classical meson
fields, time reversal symmetry, no-sea contribution, etc. Pairing is introduced
under the BCS approximation. Since the nuclear density as a function of radius
is very important in our calculation, we have solved the equations in 
co-ordinate space. The strength of the zero range pairing force is taken as 
300 MeV-fm for both protons and neutrons. For the odd mass nuclei also, we 
have followed the same procedure though the time reversal symmetry is not 
an exact symmetry for them. The main aim of the present work is to
calculate density and use them to predict the half lives for $\alpha$ decay.
We expect the
density for the odd mass nuclei calculated with time reversal symmetry 
to be nearly identical to the results when explicit breaking of symmetry is 
taken into account.

Rashdan\cite{Rashdan1} has also used the RMF Lagrangian to consistently 
calculate nucleus nucleus potential and obtained the cross sections for 
elastic scattering in halo nuclei. That work reproduces the work better than
the DDM3Y force in case of scattering of the halo nucleus $^{11}$Li.
However, we expect that in the nuclei that we have studied, the 
conventional DDM3Y potential may be adequate to calculate the
$\alpha$-daughter nucleus potential. 
Indeed, Rashdan\cite{Rashdan1} has pointed out that the 
optical model potential had to be strongly reduced to explain the measured 
angular distribution but such a large reduction cannot explain the 
total cross section. However, we have observed that in our work, there was 
no renormalization except the effect that
it may have on the spectroscopic factor. Thus we have
used the DDM3Y interaction in our work.

The microscopic density dependent M3Y interaction (DDM3Y) was obtained 
from a finite range nucleon interaction by introducing a density dependent 
factor. This class of interactions has been employed widely in the study of
nucleon-nucleus as well as nucleus-nucleus scattering, calculation of proton
radioactivity, etc.
In this work,
we have employed the interaction DDM3Y1 which has an 
exponential density dependence
\bea
v(r,\rho_1,\rho_2,E)=C(1+\alpha\exp{(-\beta(\rho_1+\rho_2)})
(1-0.002E)u^{M3Y}(r)\eea
used in Ref. \cite{Khoa} to study $\alpha$-nucleus scattering. 
Here $\rho_1$ and $\rho_2$ are the densities of 
the $\alpha$-particle and the daughter nucleus, respectively and $E$ is the 
energy per nucleon
of the $\alpha$-particle in MeV. 
It uses the direct
M3Y potential $u^{M3Y}(r)$ based on the $G$-matrix elements of the
Reid\cite{Reid} NN potential. The weak energy dependence was introduced\cite{Khoa1} to
reproduce the empirical energy dependence of the optical potential.
The 
parameters used have been assigned the standard values {\em viz.} $C=0.2845$, $\alpha=3.6391$ 
and $\beta=2.9605$ fm$^2$ in this work. 
This interaction has been  folded 
with the theoretical densities of $\alpha$ particle and the daughter nucleus
in their ground states using the code DFPOT\cite{dfpot} to obtain the 
interaction between them. 

Once the $\alpha$-nucleus interaction has been obtained, the barrier tunneling 
probability for the $\alpha$ particle has been calculated in the WKB 
approximation. The assault frequency has been calculated from the decay energy 
following Gambhir 
{\em et al}\cite{gambhir}. All the lifetime values calculated are for $l=0$
decays, {i.e.} assuming no centrifugal barrier. For odd mass nuclei, it is 
possible that some of the decays 
involve non-zero $l$ values. However, as no experimental
evidence is available for the spin-parity of the levels involved in the decay, 
we have not included the centrifugal barrier.

We have assumed the nuclei studied to be spherical in shape. Other relativistic
structure calculations have also suggested that nuclei in the vicinity of 
$^{286}$114 are spherical in shape as 
$N\sim172,\:Z\sim~114$ behave like a closed core. Our results for binding energy
and $Q$-value of $\alpha$-decay for the different chains are presented in
TABLE \ref{tab1}. No experimental binding energy value is available for any 
of the nuclei studied. Except for the decay of $^{283}$113 and $^{283}$112, the
$Q$-values are close to experiment. For the above two nuclei, the $Q$-value
predictions are poor because the daughter nuclei in the two cases,  $^{279}$111
and $^{279}$110 respectively, have been
assumed to be spherical. In reality, they are more likely to be deformed and 
hence the binding energies of the daughters are larger than predicted by 
spherical calculation. 
 
\begin{figure}[h]

\includegraphics{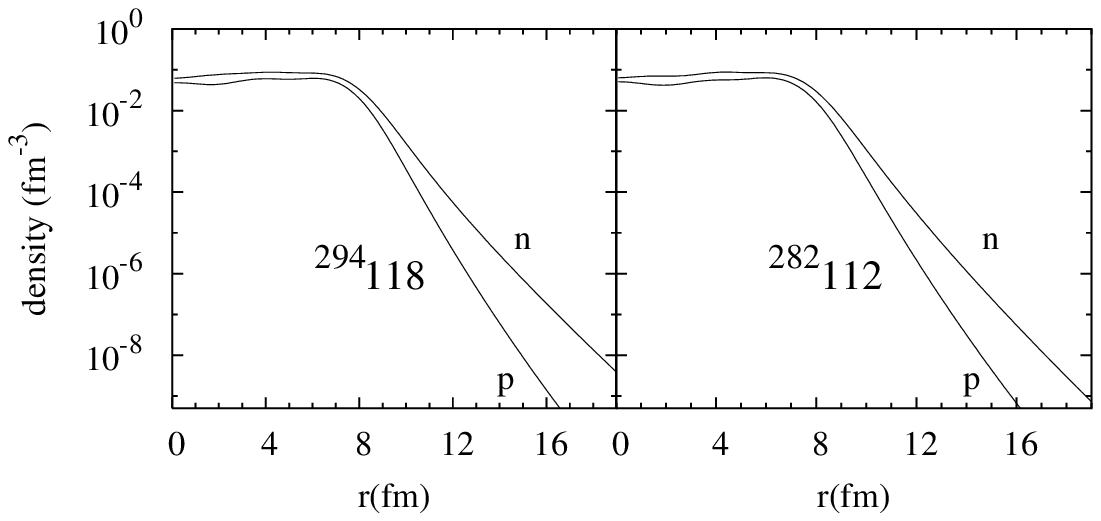}
\caption{Calculated nucleon densities in $^{294}$118 and $^{282}$112.
Here n(p) refer to neutron(proton) densities, respectively.\label{figD1}}

\includegraphics{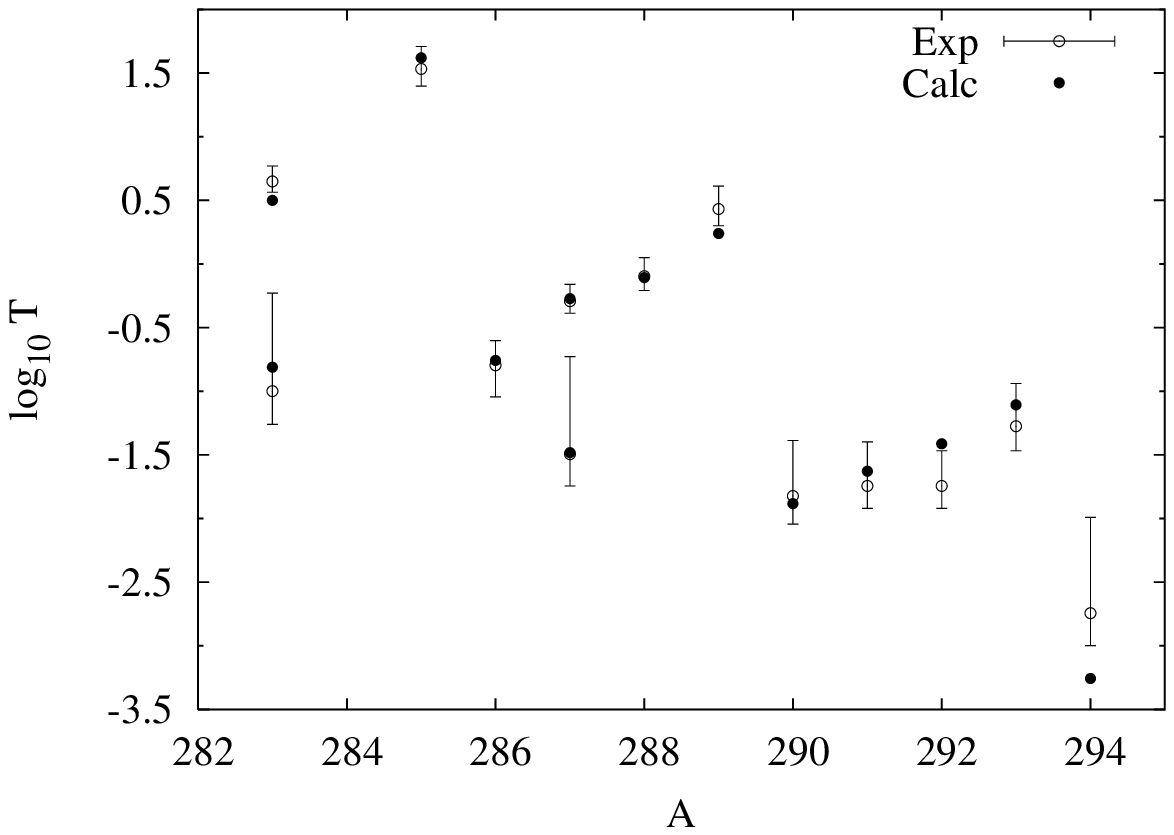}
\caption{Calculated and experimental half life values in superheavy nuclei
\label{figP1}}
\end{figure}

The density as a function of radius is very important in calculating the
$\alpha$-nucleus potential. In FIG. \ref{figD1}, the proton and neutron densities 
in two nuclei, $^{294}$118 and $^{282}$112 have been plotted. The densities for 
other nuclei studied in this work also follow the same pattern.

A small change in $Q$-value can lead to an order of magnitude change in the 
estimates of life time and theoretically calculated $Q$-values do not achieve 
such high accuracy. Following the usual practice, the $Q$-values (and the decay 
energies) have been taken from experiment in the present work. 
 
The spectroscopic factor in $\alpha$-decay  was introduced to incorporate
the preformation probability. It mainly contains the nuclear structure effects, 
and may be thought as the overlap between the actual ground state configuration 
of the parent and the configuration described by one $\alpha$-particle coupled 
to the ground state of the daughter. Obviously, it is expected to be much less 
than unity as there are contributions from many other configurations other than 
the one mentioned above. As we have considered only a small mass region, 
$283\le A\le 294$ for the parent nuclei, we do not expect the spectroscopic 
factor to vary to any large extent.
 In the present work, we have not calculated the
spectroscopic factor theoretically but have taken a constant value 
$1.4\times10^{-2}$ for 
all the decays from a fit of the half life values. This number is smaller than 
the values assumed usually. 
However, Mohr\cite{Mohr1} have shown that the spectroscopic factors
are considerably small, and more so in the region 
above $A=280$. Applying this value for the spectroscopic factor in all the 
decays, we obtain the half life values.

Our results for half life values are shown in FIG. \ref{figP1}. One can see that 
in most cases, the agreement is quite satisfactory. Only in the case of the 
decay of $^{294}$118 do we have a significant departure. Of course, one must 
remember that the errors in the measured values are large because of the 
experimental difficulties and consequent poor statistics. The errors shown 
correspond to $1\sigma$ values. The excellent agreement over a range
over four orders of magnitude shows that the assumption that the nuclei under 
investigation are spherical in shape is fairly correct. It also shows that 
the densities are fairly well reproduced in the present calculation with the 
force FSU Gold.

To summarize, $\alpha$ decay half lives in  SHN  with $A>282$ have been 
calculated in SAFM. The nuclear binding energy, $Q$-value and density have been
obtained from RMF approach in co-ordinate space using the force FSU Gold and a 
zero range pairing 
interaction. Alpha nucleus potential has been constructed with the DDM3Y1 
interaction, which uses an exponential density dependence, in the
double folding model with the densities of the daughter nucleus and the
$\alpha$ particle. Lifetimes of $\alpha$ decay have been calculated for tunneling of
the $\alpha$ particle through the potential barrier in the WKB approach
 and assuming a constant preformation probability. The resulting values agree 
with experimental
measurements.


This work was carried out with financial assistance of the
Board of Research in Nuclear Sciences, Department of Atomic Energy (Sanction
No. 2005/37/7/BRNS). Discussion with Subinit Roy is gratefully acknowledged.

\newpage

\begin{table}
\begin{center}
\caption{Binding energy and $\alpha$-decay $Q$-values for the nuclei studied
\label{tab1}}
\begin{tabular}{cccc}\hline
Nucleus & B.E./A & \multicolumn{2}{c}{$Q_\alpha(MeV)$} \\
&(MeV)&Theo. & Expt.\\\hline
$^{294}$118& 7.107&11.453&11.81\\
$^{290}$116& 7.146& 10.979& 11.00\\
$^{286}$114& 7.186&  9.830& 10.345\\
$^{282}$112& 7.222\\
\hline                                                                          
$^{292}$116& 7.140& 10.673& 10.80\\
$^{288}$114& 7.178&  9.287& 10.09\\
$^{284}$112& 7.212\\
\hline                                                                          
$^{287}$115& 7.170& 10.535& 10.74\\
$^{283}$113& 7.208&  8.226& 10.26\\
$^{279}$111& 7.240\\
\hline                                                                         
$^{293}$116& 7.136& 10.543& 10.67\\
$^{289}$114& 7.174&  9.047&  9.96\\
$^{285}$112& 7.207&  8.745& 9.29\\
$^{281}$110& 7.240\\
\hline                                                                  
$^{291}$116& 7.143& 10.805& 10.89\\
$^{287}$114& 7.182&  9.568& 10.16\\
$^{283}$112& 7.217&  6.597& 9.76\\
$^{279}$110& 7.243\\\hline
\end{tabular}
\end{center}
\end{table}

\end{document}